\documentclass[prb,aps,twocolumn,amsmath,amssymb,floatfix,showpacs,
superscriptaddress]{revtex4}

\usepackage[dvips]{graphics}
\usepackage{color}

\begin{document}

\title{Magneto-transport in a quantum network: Evidence of a mesoscopic 
switch}

\author{Srilekha Saha}

\affiliation{Theoretical Condensed Matter Physics Division, Saha
Institute of Nuclear Physics, Sector-I, Block-AF, Bidhannagar,
Kolkata-700 064, India}

\author{Santanu K. Maiti}

\email{santanu@post.tau.ac.il}

\affiliation{School of Chemistry, Tel Aviv University, Ramat-Aviv,
Tel Aviv-69978, Israel}

\author{S. N. Karmakar}

\affiliation{Theoretical Condensed Matter Physics Division, Saha
Institute of Nuclear Physics, Sector-I, Block-AF, Bidhannagar,
Kolkata-700 064, India}

\begin{abstract}

We investigate magneto-transport properties of a $\theta$ shaped three-arm 
mesoscopic ring where the upper and lower sub-rings are threaded by 
Aharonov-Bohm fluxes $\phi_1$ and $\phi_2$, respectively, within a 
non-interacting electron picture. A discrete lattice model is used to 
describe the quantum network in which two outer arms are subjected to 
binary alloy lattices while the middle arm contains identical atomic 
sites. It is observed that the presence of the middle arm provides 
localized states within the band of extended regions and lead to the 
possibility of switching action from a high conducting state to a low 
conducting one and vice versa. This behavior is justified by studying 
persistent current in the network. Both the total current and individual 
currents in three separate branches are computed by using second-quantized 
formalism and our idea can be utilized to study magnetic response in any 
complicated quantum network. The nature of localized eigenstates are also 
investigated from probability amplitudes at different sites of the quantum 
device.

\end{abstract}


\maketitle

\section{Introduction}

Theoretical and experimental investigations in low-dimensional systems 
lead to the opportunity of visualizing various novel quantum mechanical 
effects~\cite{Jaya1,Jaya2} in a tunable way. Persistent current being one 
such exotic quantum mechanical phenomenon observed in normal metal 
mesoscopic rings and nanotubes pierced by Aharonov-Bohm (AB) flux 
\begin{figure}[ht]
{\centering \resizebox*{2.75cm}{4cm}{\includegraphics{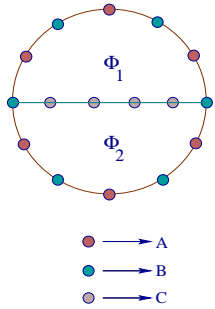}}\par}
\caption{(Color online). Schematic view of a quantum network where the
upper and lower sub-rings threaded by AB fluxes $\phi_1$ and $\phi_2$, 
respectively. Both the upper and lower arms are subjected to binary alloy
lattices, while the middle arm contains identical lattice sites. The filled 
colored circles correspond to the positions of the atomic sites.}
\label{ring}
\end{figure}
$\phi$. Prior to its experimental evidence, the possibility of a non-decaying 
current in normal metal rings was first predicted by B\"{u}ttiker, Imry and 
Landauer~\cite{Butti} in a pioneering work, and, in the sub-sequent years 
theoretical attempts were made~\cite{gefen,altshu,schmid,maiti1,san5,bell,
chen,wu,li1,li2,san1,san3,san4,san6} to understand the actual mechanism 
behind it. The experimental realization of this phenomenon of non-decaying 
current in metallic rings/cylinders has been established quite in late. 
It has been first examined by Levy {\em et al.}~\cite{levy} and later many 
other experiments~\cite{chand,mailly,blu} have confirmed the existence of 
non-dissipative currents in such quantum systems. 

Although the studies involving simple mesoscopic rings have already generated
a wealth of literature there is still need to look deeper into the problem
to address several important issues those have not yet been explored, as for
example the understanding of the behavior of persistent current in multiply 
connected quantum network, specially in presence of disorder. It is well 
known that in presence of random site-diagonal disorder in an one-dimensional 
($1$D) mesoscopic ring all the energy levels are localized~\cite{ander}, and 
accordingly, the persistent current gets reduced enormously in presence of 
disorder compared to that of an ordered ring. But there are some $1$D 
disordered systems which support extended eigenstates along with the 
localized energy levels, and these materials may provide several interesting 
issues, mainly to provide a localization to delocalization transition and 
vice versa. For example, in a pioneering work Dunlap 
{\em et al.}~\cite{Phillip1} have shown that even in 1D disordered systems 
extended eigenstates are possible for certain kind of topological 
correlations among the atoms. They have proposed that any physical system 
which can be described by the random dimer model should exhibit the 
transmission resonances and a huge enhancement in the transmission takes 
place when the Fermi level coincides with the unscattered states. In a 
consecutive year Wu {\em et al.}~\cite{Phillip2} have argued that the 
random dimer model can also be used to explain the insulator to metal 
transition in polyaniline as a result of the movement of the Fermi level 
to extended region. Later, several other works~\cite{liu,hu1,hu2,arun1,
arun2} have also been carried out in such type of materials to exhibit 
many important physical results.

The existence of localized energy eigenstates together with 
the extended states in a simple ring geometry has been explored in some 
recent works by Jiang {\em et al.}~\cite{hu1,hu2}. They have analyzed the
nature of these states by evaluating persistent current and the wave 
amplitudes at different sites of the ring. In these systems localized 
states appear by virtue of disorder. But, in our present work we make an
attempt to establish localized eigenstates, even in the absence of disorder, 
along with extended states simply by considering the effect of topology
of the system. To the best of our knowledge, this behavior has not been 
addressed earlier in the literature. Here we consider a three-arm 
mesoscopic ring in which two outer arms are subjected to binary alloy 
lattices and the middle one 
contains identical lattice sites, and, we show that due to the presence 
of the middle arm quasi-localized energy eigenstates are observed within 
the band of extend regions. It leads to the possibility of getting switching 
action from a high conducting state to a low one and vice versa as a result
 of the movement of the Fermi level. We illustrate this behavior by studying 
persistent current in the quantum network and explore the nature of energy 
eigenstates in terms of the probability amplitude in different lattice sites 
of the geometry. Our present analysis can be utilized to study magnetic 
response in any complicated quantum network and we believe that this work 
offers an excellent opportunity to study the simultaneous effects of 
topology and the magnetic fields threaded by two sub-rings in our three-arm 
ring system.

With an introduction in Section I we organize the paper as follows. In 
Section II, first we present the model, then describe the theoretical 
formalism which include the Hamiltonian and the formulation of persistent 
currents in individual branches of the network. In Section III we analyze 
the results and finally in Section IV we draw our conclusions.

\section{Model and Theoretical Formulation}

\subsection{The model and the Hamiltonian}

Let us refer to Fig.~\ref{ring}. A three-arm mesoscopic ring where the upper 
and lower sub-rings are threaded by AB fluxes $\phi_1$ and $\phi_2$,
respectively. The outer arms are subjected to binary alloy lattices (consisting
of A and B types of atoms) and the middle arm contains identical lattice sites
(atomic sites labeled by C) except those on the boundaries. The filled colored 
circles correspond to the positions of the atomic sites. Within a tight-binding 
framework the Hamiltonian for such a network reads as,
\begin{eqnarray}
H & = & \sum_{j} \epsilon_{j} c_j^\dag 
c_{j} + t \sum_{j} \left(c_{j}^{\dag} c_{j+1} 
e^{-i\theta_1} + h.c.\right) \nonumber \\
& + &\sum_{l} \epsilon_{l} c_{l}^\dag 
c_{l} + v \sum_{l} \left(c_{l}^{\dag} c_{l+1} 
e^{-i\theta_2} + h.c.\right) 
\label{eq4}
\end{eqnarray}
where, $\epsilon_j$ represents the site energy for the outer arms, while 
for the middle arm it is assigned by $\epsilon_l$. In the outer ring 
$\epsilon_j=\epsilon_A$ or $\epsilon_B$ alternately so that it forms a 
binary alloy. On the other hand, $\epsilon_l=\epsilon_C$ for the atomic 
sites those are referred by C atoms. $t$ and $v$ are the nearest-neighbor 
hopping integrals in the outer and middle arms, respectively. Due to the 
presence of magnetic fluxes $\phi_1$ and $\phi_2$ in two sub-rings, phase 
factors $\theta_1$ and $\theta_2$ appears into the Hamiltonian. They 
are expressed as follows: $\theta_1=2\pi(\phi_1+\phi_2)/(N_U+N_L)$ and
$\theta_2=2\pi(\phi_1-\phi_2)/2 N_M$. Here the fluxes are measured in units 
of the elementary flux-quantum $\phi_0$ ($=ch/e$), and, $N_U$, $N_M$ and 
$N_L$ represent the total number of single bonds (each single bond is formed 
by connecting two neighboring lattice sites) in the upper, middle and lower 
arms, respectively. It reveals that $N_U+N_M+N_L-1$ number of atomic sites 
in the quantum network. $c_j^{\dag}$ ($c_j$) corresponds to the creation 
(annihilation) operator of an electron at the $j~\mbox{th}$ site, and, a 
similar definition also goes for the atomic sites $l$.    

\subsection{Calculation of persistent current}

In the second quantized notation the general expression of charge current 
operator is in the form~\cite{san2},
\begin{equation}
I=\frac{2\pi i e \alpha}{L}\sum_n \left(c_n^{\dag}c_{n+1}-
c_{n+1}^{\dag}c_n\right).
\end{equation}
Here, $L$ is the length of the arm in which we are interested to calculate 
the current and $\alpha$ represents the nearest-neighbor hopping strength. 
The nearest-neighbor hopping strength ($\alpha$) is equal to $t$ for the 
outer arms, while for the middle arm it becomes identical to $v$. Therefore, 
for a particular eigenstate $|\psi_k\rangle$ the persistent current becomes, 
$I^k=\langle \psi_k|I|\psi_k \rangle$, where 
$|\psi_k\rangle=\sum_p a_p^k |p\rangle$. Here $|p\rangle$'s are the Wannier 
states and $a_p^k$'s are the corresponding coefficients.

Following the above relations, now we can write down the expressions for 
persistent currents in the individual branches for a given eigenstate 
$|\psi_k\rangle$. They are as follows.\\
\vskip 0.1cm
\noindent
For the upper arm:
\begin{equation}
I_U^k=\frac{2\pi i e t}{N_U+N_L} \sum_j \left(a_j^{k*} a_{j+1}^k 
e^{-i \theta_1} - h.c.\right)
\label{eq44}
\end{equation}
where, summation over $j$ spans from $1$ to $N_U$. 
\vskip 0.1cm
\noindent
In the case of middle-arm:
\begin{equation}
I_M^k=\frac{\pi i e v}{N_M} \sum_l \left(a_l^{k*} a_{l+1}^k 
e^{-i \theta_2} - h.c.\right)
\label{eq5}
\end{equation}
here, the net contribution comes from $N_M$ bonds.
\vskip 0.1cm
\noindent
Finally, for the case of lower arm:
\begin{equation}
I_L^k=\frac{2\pi i e t}{N_U+N_L} \sum_j \left(a_j^{k*} a_{j+1}^k 
e^{-i \theta_1} - h.c.\right)
\label{eq6}
\end{equation}
In this case the net contribution comes from the lower bonds. The lattice
constant $a$ is set equal to $1$.

At absolute zero temperature ($T=0\,$K), the net persistent current in any 
branch of the quantum network for a particular electron filling can be 
obtained by taking sum of the individual contributions from the energy 
levels with energies less than or equal to Fermi energy $E_F$. Therefore, 
for $N_e$ electron system total persistent in any branch becomes 
$I_{\beta}=\sum_k^{N_e} I^k_{\beta}$, where $\beta=U$, $M$ and $L$, for 
the upper, middle and lower arms, respectively. Once $I_U$, $I_M$ and $I_L$ 
are known, the net persistent current for the full network can be easily 
obtained simply adding the contributions of the individual arms, and hence 
the total current is given by $I_T=I_U+I_M+I_L$.

The net persistent current ($I_T$) can also be determined in some other ways
as available in the literature. Most probably the easiest way of calculating 
persistent current is to take first order derivative of ground state energy 
with respect to magnetic flux~\cite{maiti1,san5}. However in this method it 
is not possible to find the distribution of persistent current in individual
arms of the network with a high degree of accuracy. On the other hand in our 
present scheme, the so-called second-quantized approach, there are certain 
advantages compared to other available procedures. Firstly, we can easily
calculate persistent current in any branch of a network. Secondly, the 
determination of individual responses in separate arms provides much deeper
insight to the actual mechanism of electron transport in a transparent way.

In the present work we investigate all the essential feature of 
magneto-transport at absolute zero temperature and choose the units where 
$c=e=h=1$. Throughout the numerical work we set $t=v=-1$ and measure the 
energy scale in unit of $t$.

\section{Numerical results and discussion}

\subsection{Quantum network with \mbox{\boldmath $\epsilon_A=\epsilon_B=0$}}

We first start with a perfect quantum system where $\epsilon_A$ and 
$\epsilon_B$ and $\epsilon_C$ are all identical to each other and we set 
$\epsilon_A=\epsilon_B \epsilon_C=0$. To have a 
\begin{figure}[ht]
{\centering \resizebox*{8cm}{7cm}{\includegraphics{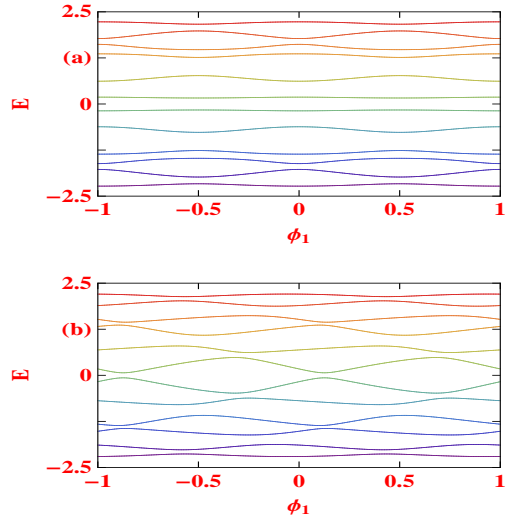}}\par}
\caption{(Color online). Energy levels as a function of flux $\phi_1$ for
a three-arm ring with $\epsilon_A=\epsilon_B=0$ considering $N_U+N_L=10$ 
and $N_M=3$, where (a) and (b) correspond to $\phi_2=0$ and $\phi_0/4$,
respectively.}
\label{energy1}
\end{figure}
clear idea about the magnetic response of the model quantum system, first we 
illustrate the behavior of energy spectra as a function of flux $\phi_1$ for 
different values of flux $\phi_2$ threaded by the lower sub-ring. The results 
are presented in Fig.~\ref{energy1}, where (a) and (b) correspond to $\phi_2=0$ 
and $\phi_0/4$, respectively. In the absence of flux $\phi_2$, energy levels 
near the edges of the spectrum become more dispersive than those lying in the 
central region (see Fig.~\ref{energy1}(a)) and  near the center of the 
spectrum the energy levels are almost non-dispersive with respect to flux 
$\phi_1$. This feature implies that the persistent current amplitude becomes 
highly sensitive to the electron feeling i.e., the Fermi energy $E_F$ of the 
system, since the current is directly proportional to the slope of the energy 
levels~\cite{maiti1}. The situation becomes much more interesting when we add 
a magnetic flux in the lower sub-ring. Here, the energy levels near the 
central region of the spectrum becomes more dispersive in nature 
\begin{figure}[ht] 
{\centering \resizebox*{8.7cm}{9cm}{\includegraphics{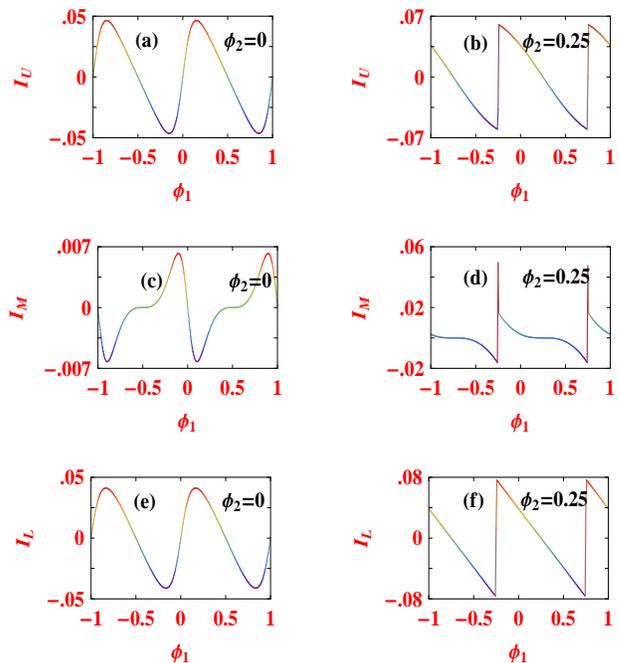}}\par}
\caption{(Color online). Persistent current in different arms as a 
function of $\phi_1$ for a three-arm ring with $\epsilon_A=\epsilon_B=0$ 
in the half-filled band case considering $N_U+N_L=60$ and $N_M=25$.}
\label{current1}
\end{figure}
than the energy levels near the edges (Fig.~\ref{energy1}(b)), and it
increases gradually with flux $\phi_2$, which gives a possibility of getting
higher current amplitude with increasing the total number of electrons $N_e$ 
in the system. A similar kind of energy spectrum is also observed if we plot 
the energy levels as a function of flux $\phi_2$ instead of $\phi_1$, 
keeping $\phi_1$ as a constant. All these energy levels exhibit $\phi_0$
($=1$, in our choice of units $c=e=h=1$) flux-quantum periodicity. Thus, 
for such a simple quantum network persistent current amplitude might be 
regulated for a particular filling simply by tuning the magnetic flux 
threaded by anyone of two such sub-rings, and, its detailed descriptions
are available in the sub-sequent parts.

In Fig.~\ref{current1} we present the variation of persistent current in
individual arms of the three-arm quantum network as a function of flux 
$\phi_1$ for some fixed values of $\phi_2$. The panels from the top 
correspond to the results for the upper, middle and lower arms, respectively, 
\begin{figure}[ht]
{\centering \resizebox*{6.8cm}{9cm}{\includegraphics{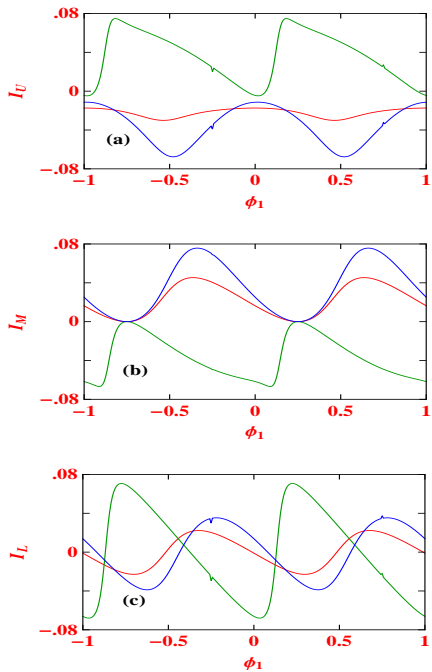}}\par}
\caption{(Color online). Persistent current in different arms as a function
of $\phi_1$ for a three-arm ring with $\epsilon_A=\epsilon_B=0$ considering 
$N_U+N_L=40$ and $N_M=17$. The red, green and blue curves correspond to 
$N_e=10$, $15$ and $20$, respectively. For all these spectra $\phi_2$ is 
set at $\phi_0/4$.}
\label{currentne}
\end{figure}
and in all these cases the current is determined for the half-filled band 
case i.e., $N_e=42$. The left column represents the current for $\phi_2=0$,
and the right column gives the current when $\phi_2$ is fixed at 
$\phi_0/4$. From the spectra we notice that in some cases current shows
continuous like behavior while in some other cases it exhibits saw-tooth
like nature as a function of flux $\phi_1$ threaded by the upper sub-ring.
This saw-tooth or continuous like feature solely depends on the behavior 
of the ground state energy for a particular filling ($N_e$). It is to be
noted that in a conventional ordered AB ring we always get saw-tooth like 
behavior of persistent current irrespective of the filling of the 
electrons~\cite{maiti1}.
In the saw-tooth variation a sudden change in direction of persistent current 
takes place across a particular value of magnetic flux which corresponds to 
a phase reversal from the diamagnetic nature to the paramagnetic one or vice 
versa. In our three-arm geometry we also observe that though the current in 
the upper arm or in the lower arm is not so sensitive to the flux $\phi_2$,
but the current amplitude in the middle arm changes remarkably, even an order 
of magnitude, in presence of flux $\phi_2$, which leads to a net larger 
current since the total current is obtained by adding the contributions from
the individual arms. 

To explore the filling dependent behavior of persistent current, in 
Fig.~\ref{currentne} we display persistent currents for three different 
arms as a function of flux $\phi_1$ for a typical value of $\phi_2$.
Here, $\phi_2$ is set at $\phi_0/4$. The red, green and blue lines 
represent the currents for $N_e=10$, $15$ and $20$, respectively. The
current in different arms shows quite a complex structure which strongly
depends on the electron filling as well as magnetic flux $\phi_2$. In all
these cases persistent current provides $\phi_0$ flux-quantum periodicity,
like a traditional single-channel mesoscopic ring or a multi-channel
cylinder. 

\subsection{Quantum network with \mbox{\boldmath $\epsilon_A\ne\epsilon_B$}}

Now we focus our attention to the geometry where site energies in the outer
arms are no longer identical to each other i.e., $\epsilon_A \ne \epsilon_B$.
In this case energy spectrum gets modified significantly compared to the
\begin{figure}[ht]
{\centering \resizebox*{8cm}{7cm}{\includegraphics{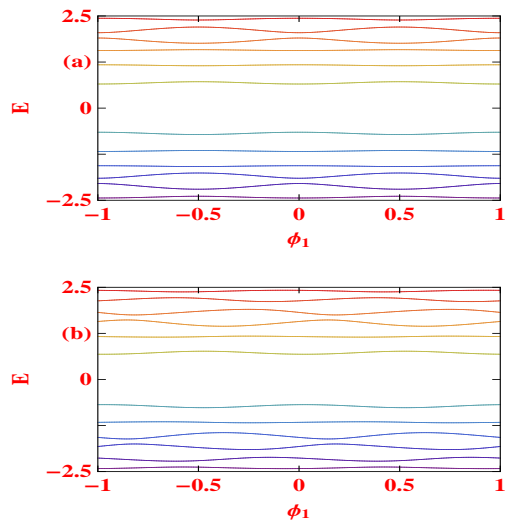}}\par}
\caption{(Color online). Energy levels as a function of flux $\phi_1$ for
a three-arm ring with $\epsilon_A=-\epsilon_B=1$ considering $N_U+N_L=10$ 
and $N_M=3$, where (a) and (b) correspond to $\phi_2=0$ and $\phi_0/4$,
respectively.}
\label{energy2}
\end{figure}
previous one where site energies are uniform ($\epsilon_A=\epsilon_B$). To
illustrate it in Fig.~\ref{energy2} we plot the energy-flux characteristics
for a three-arm quantum network considering $\epsilon_A=-\epsilon_B=1$, 
where (a) and (b) correspond to the identical meaning as given in
Fig.~\ref{energy1}. Since the upper and lower arms of the network are 
subjected to the binary alloy lattices we get two sets of discrete energy 
levels spaced by a finite gap around $E=0$ (see Fig.~\ref{energy2}).
Quite interestingly we see that the energy levels near the two extreme 
edges of the spectrum are more dispersive in nature than those situated 
along the inner region. With increasing the difference in site energies 
($|\epsilon_A-\epsilon_B|$), we get more less dispersive energy levels in 
the inner region and for large enough value of $|\epsilon_A-\epsilon_B|$ 
these levels become almost non-dispersive and they practically contribute 
nothing to the current. Thus, for such a system a mixture of quasi-extended 
and quasi-localized energy levels are found out and it can provide a very 
large or almost zero current depending on the electron filling. For a very 
\begin{figure}[ht]
{\centering \resizebox*{8.5cm}{6cm}{\includegraphics{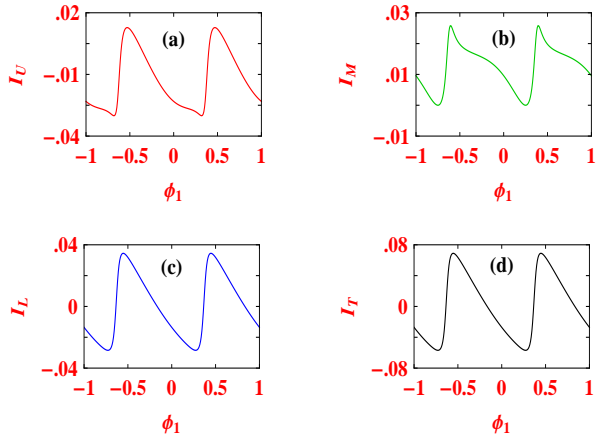}}\par}
\caption{(Color online). Current-flux characteristics of a three-arm ring 
with $\epsilon_A=-\epsilon_B=1$ considering $N_U+N_L=60$, $N_M=25$ and
$\phi_2=\phi_0/4$ in the quarter-filled ($N_e=21$) band case, where (a)-(d) 
correspond to the currents in the upper, middle and lower arms and in the 
full system, respectively.}
\label{current2}
\end{figure}
\begin{figure}[ht]
{\centering \resizebox*{8.5cm}{6cm}{\includegraphics{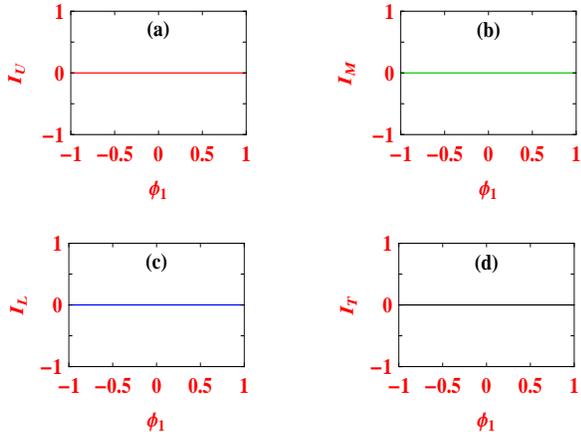}}\par}
\caption{(Color online). Current-flux characteristics of a three-arm ring 
with $\epsilon_A=-\epsilon_B=1$ in the half-filled ($N_e=42$) band case for 
the same parameter values used in Fig.~\ref{current2}.}
\label{current3}
\end{figure}
large system size, the energy separation between two successive levels in 
each set of discrete energy levels gets reduced and we get two quasi-band 
of energies separated by a finite gap, where the gap is controlled by the 
parameter values. It is important to note that, unlike the previous one 
(Fig.~\ref{energy1}), for this case the energy spectrum is not so sensitive 
to flux $\phi_2$ (Fig.~\ref{energy2}). The presence of C-type of atoms in 
the middle arm which divides the binary alloy ring into two sub-rings is 
responsible for the existence of quasi-localized energy levels near the 
inside edges of two quasi-band of energies. Thus we get more non-dispersive 
energy levels with increasing the length of the middle arm.

The existence of nearly extended and localized states becomes much more
clearly visible from our current-flux spectra. As illustrative example,
in Fig.~\ref{current2} we display the variation of persistent current in
\begin{figure}[ht]
{\centering \resizebox*{7cm}{4cm}{\includegraphics{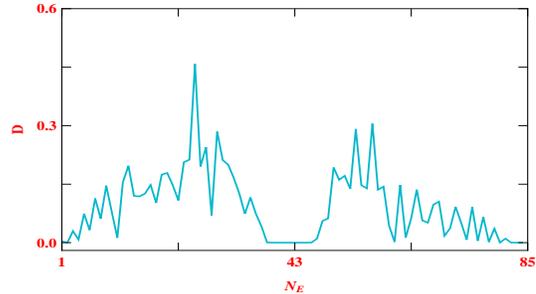}}\par}
\caption{(Color online). Charge stiffness constant ($D$) 
as a function of electron filling ($N_e$) for a three-arm ring with 
$\epsilon_A= - \epsilon_B =1$ considering $N_U + N_L =60$, $N_M=25$ and
$\phi_2=\phi_0/4$.}
\label{drude}
\end{figure}
\begin{figure}[ht]
{\centering \resizebox*{8cm}{7cm}{\includegraphics{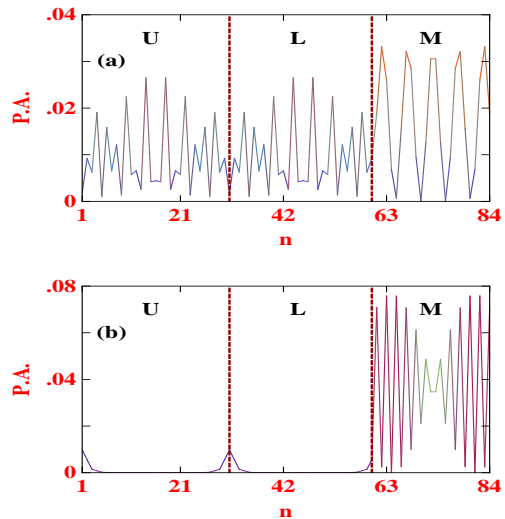}}\par}
\caption{(Color online). Probability amplitude (P.A.) as a function of 
site index ($n$) for a three-arm ring with $\epsilon_A=-\epsilon_B=1$ 
considering $N_U+N_L=60$ and $N_M=25$ when $\phi_1$ and $\phi_2$ are set 
at $\phi_0/4$. (a) and (b) correspond to the results of $21$-st and $42$-nd
eigenstates, respectively.}
\label{probability}
\end{figure}
individual arms including the total current of a three-arm ring considering
$\epsilon_A=-\epsilon_B=1$ for the quarter-filled ($N_e=21$) band case.
The flux $\phi_2$ is set equal to $\phi_0/4$. From the spectra it is 
clearly observed that the current in each arm provides a non-zero value
(Figs.~\ref{current2}(a)-(c)), and accordingly, the system supports a 
finite current as shown in Fig.~\ref{current2}(d). The situation becomes
completely opposite when the filling factor is changed. Quite remarkably
we notice that persistent current almost vanishes in three separate branches
which provides almost vanishing net current in the half-filled band case.
The results are illustrated in Fig.~\ref{current3}, where (a)-(d) correspond
to the identical meaning as in Fig.~\ref{current2}. The vanishing nature
at half-filling and the non-vanishing behavior of current when the system 
is quarterly filled can be easily understood from the following argument.
The total current in any branch or in the complete system mainly depends 
on the contributions coming from the higher occupied energy levels, while
the contributions from the other occupied energy levels cancel with each 
other. Therefore, for the quarter-filled band case, the net contribution
comes from the energy levels which are quasi-extended in nature and a 
non-zero current appears. On the other hand, for the half-filled band 
case, the net contribution arises from the levels those are almost 
localized, and hence, nearly vanishing current is obtained. Thus, we
can emphasize that the three-arm ring leads to a possibility of getting 
high-amplitude to low-amplitude (almost zero) persistent current simply
by tuning the filling factor $N_e$ i.e., the Fermi energy $E_F$, and,
hence the network can be used as a mesoscopic switch.

The high-conducting to low-conducting switching action 
with the change of electron filling $N_e$ in our topology can also be very 
well explained from the spectrum given in Fig.~\ref{drude}, where we 
measure the conducting nature by calculating charge stiffness constant, 
the so-called Drude weight ($D$), in accordance with the idea originally 
put forward by Kohn~\cite{kohn}. The Drude weight for the system can be 
easily determined by taking the second order derivative of the ground state
energy for a particular filling with respect to flux $\phi_1$ 
($\phi_1 \rightarrow 0$) threaded by the ring~\cite{kohn,san3,san33}. 
Kohn has shown that $D$ decays exponentially to zero for an insulating 
system, while it becomes finite for a conducting system. A nice feature
of the result shown in Fig.~\ref{drude} is that, the charge stiffness 
constant almost drops to zero around the half-filled region which reveals 
the insulating phase, while away from this region it ($D$) has a finite 
non-zero value that indicates a conducting nature. This feature corroborates 
the findings presented in Figs.~\ref{current2} and \ref{current3}.

To ensure the extended or localized nature of energy eigenstates, finally
we demonstrate the variation of probability amplitude (P.A.) of the
eigenstates as a function site index $n$. The probability amplitude of
getting an electron at any site $n$ for a particular eigenstate 
$|\psi_k\rangle$ is obtained from the factor $|a_n^k|^2$. Here we analyze 
the localization behavior for two different energy eigenstates, viz, 
$21$-st and $42$-nd states. For the first one the energy is located well 
inside a quasi-band, while for the other the energy is placed at the edge of 
this band. The results are given in Fig.~\ref{probability} for a three-arm ring 
with $84$ atomic sites. The red dashed lines are used to separate the three 
distinct regions of the network. In Fig.~\ref{probability}(a) we present the 
probability amplitudes of the 21st eigenstates and see that the probability 
amplitude becomes finite for any site $n$ which indicates that the energy 
eigenstate is quasi-extended. While, for the other state ($42$-nd) the 
probability amplitude almost vanishes at every site of the upper and lower 
arms of the network. Only at the atomic sites of the middle arm we have 
finite probability amplitudes. This state does not contribute anything 
to the current and we can refer the state as a localized one.

\section{Conclusion}

To summarize, in the present work we have explored the magneto-transport 
properties of a $\theta$ shaped three-arm quantum ring in the 
non-interacting electron framework. The upper and lower sub-rings of the 
network are threaded by magnetic fluxes $\phi_1$ and $\phi_2$, respectively. 
We have used a single-band tight-binding Hamiltonian to illustrate the model 
quantum system, where the outer arms are subjected to the binary alloy 
lattices and the middle arm has identical lattice sites. In the absence of 
the middle arm, all the energy eigenstates are extended, but the inclusion 
of the middle arm produces some quasi localized states within the band of 
extended states and provides a possibility of getting a high conducting 
state to the low conducting one upon the movement of the Fermi energy. 
Thus, the system can be used a mesoscopic switch. We have verified the 
switching action from high- to low-conducting state and vice versa by 
investigating the persistent current and charge stiffness constant in the 
network for different band fillings. We have numerically computed both the 
total current and the individual currents in separate branches by using 
second-quantized approach. We hope our present analysis may be helpful for 
studying magneto-transport properties in any complicated quantum network. 
Finally, we have also examined the nature of the energy eigenstates in 
terms of the probability amplitude in different sites of the geometry.

\end{document}